\documentclass{appolb}
\usepackage{graphicx}

\begin{document}
\title{The collective behavior of the partons and its influence on the jet suppression in heavy ion collisions.%
}
\author{Mais Suleymanov
\address{COMSATS University,\\Park Road Islamabad, Pakistan}
\\
{\it{mais$\_$suleymanov@comsats.edu.pk
}}
}
\maketitle
\begin{abstract}
We discuss the physical picture that a parton interaction with coherent group of partons can lead to more jet quenching effect in the hot and dense matter created by heavy ion collisions at RHIC and LHC energies.We came to this picture after analyzing the behaviour of the nuclear modification factor as a function of  $p_T$  for the charged particles produced in the most central  Pb-Pb collisions at 2.76 A TeV .
In the interval $7-50 GeV/c$ the values of the  factor as a function of $p_T$ increases almost linearly with a slope is very close to  expected one for the inverse Compton effect. Around $p_T\simeq 60$ GeV/c, a regime change occurs, which is characteristic for the phenomenon.
We propose that this similarity can be explained by the inverse Compton effect for partons, which occurs via a collective parton group formation ( through the appearance of a new string as a result of fusion of strings) and its interactions with single partons in the interval of $5 < p_T < 10 GeV/c$. In the case of a coherent collision with a parton that has a lower energy than the group, the parton can gain energy through the inverse Compton effect, resulting in its acceleration  and shifting to the region of $p_T >10$ GeV/c.
After losing a significant part of its energy new string will decay into partons with lower energies - slowed partons in the interval of $p_T < 5 GeV/c$. This enhancement in the jet quenching can be observed in the interval of $2 <p_T <20 GeV/c$.
\end{abstract}

\section{Introduction}
The most important signature for the formation of the quark-gluon plasma  in heavy-ion collisions at ultrarelativistic energies is  the suppression of hadron spectra at high $p_T$~\cite{b3}-\cite{b4}. The effect is caused by energy loss  of jet partons (via collisional and radiative interactions) - jet quenching~\cite{UrsAchiJ} - \cite{AaronJ} from early hard scattering  with a hot and dense  medium before fragmenting into hadrons.
To quantitatively extract the jet transport coefficients  were performed by comparing several jet quenching model calculations with the experimental data for the nuclear modification of single inclusive hadron production at high transverse momenta~\cite{JetCollab}. Most of these models use a formalism that treats the medium as a series of static scattering centers with the parton and radiated gluons with some energy  and frequency, traveling along eikonal trajectories\cite{AaronJ}. From the combined analysis of soft single-inclusive hadron spectra, two-particle correlations and their azimuthal dependence, one knows that the matter produced in a heavy ion collisions expands rapidly \cite{UrsAchiJ} and  show a collective behavior which is likely to be formed at an early parton stage of the space-time evolution of the produced hot and dense matter ~\cite{RHIC1} - \cite{LHC1}. Thus one can support the idea that the parton collective behavior could lead to formation of coherent parton system (for example, as a result of strings fusion~\cite{Strin1} - \cite{Strin4})
in the hot and dense medium (see paper ~\cite{suley2017} ).
Appearance of the coherent parton group can influence the jet quenching picture in the $p_T$ region where the groups were formed with high probability (in the paper~\cite{suley2017} it was shown that the boundary values of $p_T$ for the region could be  4-20 GeV/c).
In this regions partons could collide with higher energy objects - coherent parton group , and as in the case of a collision of a photon with a higher energy electron, the parton can gain energy, accelerate transiting into  higher $p_T$ region. After a significant energy loss the parton group decays into partons with lower energies  - the  slowed partons  which will be in the interval of  lower $p_T$ region. As a result of the transitions of the accelerated and slowed partons from the region of high probable parton group formation,  it can be seen that  the jet quenching  becomes more stronger in this region.

In this paper, we tried to show an experimental signature of the formation of the coherent parton group and its influence on the high $p_T$ parton suppression through of parton Inverse Compton Effect (ICE). ICE states that a photon can gain energy in a collision with a more energetic electron~\cite{ICE1}-\cite{ICE3}, but can this phenomenon occur in a collision between partons? In other words, is there a parton version of ICE?

\section{About the signature}
To get an information on the formation of the coherent parton group and  its interaction with single partons we have analysed the behaviour of the nuclear modification factor ($R_{AA}$) as a function of  $p_T$ for the charged particles produced in the most central  Pb-Pb collisions at 2.76 A TeV~\cite{b1}.

The evolution picture of the $R_{AA}$~\cite{b1} with energy from the SPS to the LHC~\cite{b3}-\cite{b4},\cite{b2} shows that the creation of  high-$p_{T} $ particles in central Pb-Pb collisions is significantly suppressed in comparison to peripheral Pb-Pb and $pp$ collisions . In the range of $p_{T} =5-10{\rm \; GeV/c}$, the suppression is stronger than before observed at the RHIC ~\cite{b3}. Beyond  $p_{T} =10{\rm \; GeV/c}$ up to 20 GeV/c the $R_{AA}$ shows a rising trend as it was shown by the data from the ALICE experiment~\cite{b4}. The CMS measurement~\cite{b1} with improved statistical precision clearly shows that this rise continues at higher $p_{T}$, approaching a suppression factor $R_{AA} \approx 0.5-0.6$ in the range of 40--100 GeV/c.

The Figure 1 shows the behaviour of the $R_{AA}$ as a function of $p_T$ for the charged particles produced in the Pb-Pb collisions  with centrality of 0-5$\%$ at energies of 2.76 A GeV\footnote{the experimental data were taken from the HEP Data:  https://hepdata.net/record/ins1088823}.  There are several trends observed as $p_T$ increases:

       - for $p_T < 2 GeV/c$  the values of $R_{AA}$ increase from ~ 0.36 to ~ 0.42;

       - for $2 <p_T <7 GeV/c$ the values of $R_{AA}$ decreases to ~ 0.15 and reaches its minimum;

       - for $7 <p_T <40-50 GeV/c$ the values of $R_{AA}$ increase from ~ 0.15 to ~0.6 and reaches its maximum.

Furthermore, at $p_T\simeq 50-60 GeV/c$, a regime change occurs and the values of $R_{AA}$ remain at 0.6. It can be easily shown that the behavior of $R_{AA}$ in the interval $7<p_T <100 GeV/c$ is similar to the behavior of the photon energy distribution under ICE \cite{ICE3} (see appendix).

\begin{figure}[htb]
\centerline{%
\includegraphics[width=12.5cm]{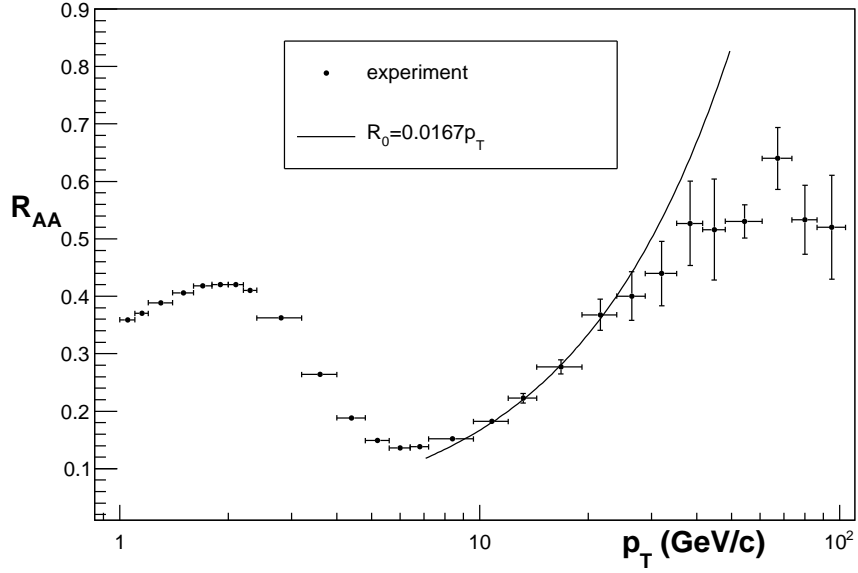}}
\caption{The behavior of $R_{AA}$ as a function of $p_T$ for charged particles produced in Pb-Pb collisions with centrality of  $0-5\%$ at energies  2.76 AGeV(see paper~\cite{b1}). The line shows the behavior of the function $R_0 = 0.0167 p_T$ for the parton ICE.}
\label{Fig:F2H}
\end{figure}

A reason for this similarity may be that in the momentum interval $5 < p_T < 10  GeV/c$ (this is the second $p_T$ region observed in the paper~\cite{suley2017}) partons collide with higher energy objects, and as in the case of a collision of a photon with a higher energy electron, parton ICE occurs. These objects can be parton groups with energy $\gamma_{group}=E_{group}/m_{group}$ ($E_{group}$ is energy of parton group and $m_{group}$ is its mass). In the case of a coherent collision of a parton with energy  $\alpha_{1part}=E_{1part}/m_{group} < \gamma_{group}$ ($E_{1part}$ is the energy of parton before collision), it can acquire energy $\alpha_{part}=E_{part}/m_{group}$ ($E_{part}$ is energy of parton after collision) and in the case of $\alpha_{part}>\alpha_{1part}$ it can accelerate due to parton ICE.

That is why we believe the observed more jet quenching in the interval of $p_T=2-60 GeV/c$ could be connected with the  formation of the coherent parton group and  its interaction with single parton which results  in the acceleration of the single partons and decay of the coherent parton group into the slow partons.

\section{Results and discussions.}

To show quantitative capacity of the physical picture under consideration we have used the analogy with photon ICE and defined the function $R_0$ to  compare with $R_{AA}$. Of course we cannot reproduce exactly quantitative contribution of the physical picture under consideration we confine oneself only with showing that the picture is acceptable.

For this we have used the approximate formulas~\cite{ICE3} are given for the energy distributions of photons $\frac{d^2N}{dtd\alpha}$   in the case of ICE when  a photon with energy $\alpha_1=E_1/m_e$   (in units of electron mass $m_e$, $E_1$ is the energy of phaton ) collides with an electron with energy $\gamma = E_e/m_e$ ($E_e$ is the electron energy) and receives energy $\alpha=E/m_e$   ($E$ photon energy):

\begin{equation}
R_{0}=\frac{\frac{d^2N}{dtd\alpha}|_{\alpha<\alpha_1}}{\frac{d^2N}{dtd\alpha}|_{\alpha=\alpha_1}}=\frac{4\gamma^2\alpha-\alpha_1}{4\gamma^2\alpha_1-\alpha}\label{eq:e1} \end{equation}
or
\begin{equation}
R_0\simeq\frac{\alpha}{\alpha_1}\label{eq:e2},
\end{equation}
for $\gamma>>1$. The $R_0$ is the energy distribution of photons ($\frac{d^2N}{dtd\alpha}$) with energy $\alpha<\alpha_1$ for  $\alpha_1 <\gamma$ (see appendix), normalized to the value of $\frac{d^2N}{dtd\alpha}$ at $\alpha =\alpha_1$.

In accordance with the physical picture, in the case of a coherent collision of a parton with energy $\alpha_{1part}<\gamma_{group}$ with a group of partons of energy $\gamma_{group}>\alpha_{1part}>>1$, parton’s final energy will be $\alpha_{part}$.
The values of $E_{1part}$ and $E_{part}$ are much larger than the parton's rest energy, so we one can assume that $E_{1part}\simeq p_{1part}$ and $E_{part}\simeq p_{part}$  ($p_{1part}$ and $p_{part}$ is 3-momentum of parton  before and after the collision). For the particles under consideration (see~\cite{b1}), the values of their pseudorapidity
are taken in the interval of $|\eta|<1$, so we can write $p_{part}\simeq p_{Tpart}$ ($p_{Tpart}$ is the transverse momentum of the parton after the collision). Therefore, for partons the expression~\ref{eq:e2} can be rewritten as:

\begin{eqnarray}
R_{0}=\frac{1}{\alpha_{1part}}\alpha_{part}\simeq \frac{1}{E_{1part}}E_{part}\simeq\nonumber\\
\simeq \frac{1}{p_{1part}}p_{part}\simeq \frac{1}{p_{1part}}p_{Tpart}\label{eq:e3}
\end{eqnarray}
To determine $p_{1part}$ from~\ref{eq:e3} we used the fact that at the point of regime change $p_{1part} = p_{part}\simeq p_{Tpart}$.
Visually, the regime change seems to be  near
$p_T\simeq p_{Tpart}\simeq 40-50  GeV/c$. To obtain a more accurate value of the $p_{1part}$  we fitted the $R_{AA}$ data with the linear function $y = ax$ in the region of  $6 <p_T <70 GeV/c$, changing the minimum and maximum values of the $p_T$ intervals to obtain the best fitting results. The Table shows three best fitting results. The fit function $y = ax$ sems to describe the $R_{AA}$ data well in the interval of $p_T = 8.4-44.8 GeV/c$, wherease after $p_T =54.4 GeV/c$ is turned on it deteriorates fast.
This means that the transition point is near the value
$p_T=55.4\pm6.4 GeV/c$ and it can be assumed that $p_{1part}\simeq 60 GeV/c$ and the value $1/p_{1part}\simeq 0.0167$  and $R_0 = 0.0167 p_T$.The solid line in the figure shows the behavior of  $R_0$
as a function of $p_T$ which well describes the $R_{AA}$ in the region $7 <p_T <50 GeV/c$. This is understandable since the slope of the line for $R_0 = 0.0167 p_T$ almost exactly coincides with the slope  of the line ($a = 0.0162\pm 0.0007$) obtained for the best fitting of the experimental data on $R_{AA}$ (see Table of the values of $a$).

The definition of $R_0$ tells us that at the regime change point the value of $R_0$ would be 1, but in the experiment the $R_{AA}$ values remain constant around 0.6 - 0.7 in the region of $p_T > 60 GeV/c$.
The difference could  be connected with two main reasons:

- we have not taken into account jet quenching and the dynamics under consideration are  unable to describe the suppression alone and it is  clear;

- due to  unavailability of a Monte Carlo simulation pocket we could not have perfect normalization for the $\frac{d^2N}{dtd\alpha}$ to compare with $R_{AA}$

As we have said above in this step we cannot reproduce exactly quantitative contribution of the parton ICE we confine oneself only with showing that the picture is acceptable.

More  important results are that: the slopes of the behaviour of $R_0$ and $R_{AA}$ are very close to each other ; there  exists the regime change  at the point $p_T\simeq 60 GeV/c$.

Finally returning back to the interval $\alpha_{part} <\alpha_{1part}$, we  noted that the similar behavior of the slope values of $R_{AA}$ behavior as a function of $p_T$ and of $R_0$ in this interval may indicate that the parton energy loss (parton suppression effect) almost does not change the slope for $R_{AA}$ in this region.

\begin{table}
\caption{\label{tab:table1}The fitting results}
\begin{tabular}{llll}\hline
The values of $p_T$(GeV/c) & $\chi^2/ndf$ & Prob. &$\;\;\;\;\;\;$ $a$ \\
values at the first   & &  &  \\
and last fittings& &  &  \\\hline
$8.4 - 38.4$&    $3.526/7$ & $0.8325$ &$0.0162\pm 0.0007$ \\
$8.4 - 44.8$&    $7.453/8$ & $0.4886$ & $0.0158\pm0.0007$ \\
$8.4 - 54.4$&    $16.85/9$ & 0.05108 & $0.0154\pm 0.0006$\\\hline
\end{tabular}
\end{table}

\section{Conclusion}

We conclude that the values of $R_{AA}$ as a function of $p_T$ in the interval $7-50 GeV/c$ increase almost linearly with a slope expected for the parton ICE. At $pT\simeq 60 GeV/c$ a regime change occurs, which is characteristic for this effect. These results indicate that partons can be accelerated by ICE in the $p_T$ region 5-10 GeV/c. We assume that this can be due to the collective effects associated with the fusion of strings and the appearance of new strings in the dense medium. In the case of a coherent collision with a parton that has a lower energy than the new string, the parton can gain energy and accelerate, transiting into $p_T> 10 GeV/c$ interval. After losing a significant part of its energy new string will decay into partons with lower energies  - slowed partons in the interval of $p_T < 5 GeV/c$. This can seem as amplification of the suppression of partons in the $2 <p_T <20 GeV/c$ interval. That is why  this  should be taken into account in the jet quenching models.

\par\bigskip

I would like to acknowledge the COMSATS University Islamabad, which provided suitable platform and all possible facilities to perform the analysis, Dr Ali Zaman and Aziza Suleymanzade for their essential help during preparing the text.

\section{Appendix}
In~\cite{ICE3} the approximate formulas are given for the energy distributions of photons $\frac{d^2N}{dtd\alpha}$   in the case of ICE. It is shown that if a photon with energy $\alpha_1=E_1/m_e$   (in units of electron mass $m_e$, $E_1$ is the energy of phaton ) collides with an electron with energy $\gamma = E_e/m_e$ ($E_e$ is the electron energy) and receives energy $\alpha=E/m_e$   ($E$ photon energy), then the energy distribution of photons after a collision can be written in the form:
\begin{equation}
\frac{d^2N}{dtd\alpha}\approx \frac{\pi r^2_0c}{2\gamma^4\alpha_1}\left(\frac{4\gamma^2\alpha}{\alpha_1}-1\right)\nonumber
\end{equation}
in the energy interval $\alpha_1/4\gamma^2\leq \alpha\leq \alpha_1$ and in the form:
\begin{eqnarray}
\frac{d^2N}{dtd\alpha}\approx \frac{2\pi r^2_0c}{\alpha_1\gamma^2}\nonumber\\
\times\left[2q^{"}\ln{q^{"}}+(1+2q^{"})(1-q^{"})
 +\frac{1}{2}
\frac{(4\alpha_1\gamma q^{"})^2}
{(1+4\alpha_1\gamma q^{"})}(1-q^{"})
\right]\nonumber
\end{eqnarray}
\begin{equation}
q^{"}=\frac{\alpha}{4\alpha_1\gamma^2(1-\frac{\alpha}{\gamma})}\nonumber
\end{equation}
in the energy interval $\alpha_1\leq\alpha\leq 4\gamma^2\alpha_1/(1+4\gamma\alpha_1)$, where $r_0$ is the classical radius of the electron. From these formulas it is clear that for values of $\alpha=\alpha_1$, a change in the regime will be observed in the behavior of $\frac{d^2N}{dtd\alpha}$  .

\end{document}